\documentclass[aps,pre,reprint,superscriptaddress,floatfix]{revtex4-2}

\usepackage{amsfonts}
\usepackage{amssymb}
\usepackage{amsmath}
\usepackage{dcolumn}  
\usepackage[T1]{fontenc}
\usepackage{silence}
\usepackage{patches}
\usepackage{hyperref}
\usepackage{cleveref}
\usepackage{makecell}
\usepackage{xcolor}
\definecolor{myred}{RGB}{255,0,0}
\definecolor{myblue}{RGB}{0,0,255}
\definecolor{mygreen}{RGB}{0,255,0}
\usepackage{siunitx}

\begin{document}

\title{Hydrodynamic Modeling of Odd Nematic Elasticity in Liquid Crystals}

\author{Zeyang Mou}
\affiliation{
Department of Physics, Hong Kong University of Science and Technology, Clear Water Bay, Kowloon, Hong Kong SAR
}

\author{Haijie Ren}
\affiliation{
Department of Physics, Hong Kong University of Science and Technology, Clear Water Bay, Kowloon, Hong Kong SAR
}

\author{Ding Xu}
\affiliation{
Department of Physics, Hong Kong University of Science and Technology, Clear Water Bay, Kowloon, Hong Kong SAR
}

\author{Igor 
S. Aranson}
\email[]{isa12@psu.edu}
\affiliation{
Departments of Biomedical Engineering, Chemistry, and
Mathematics, Pennsylvania State University, University
Park, Pennsylvania 16802
}

\author{Rui Zhang}
\email[]{ruizhang@ust.hk}
\affiliation{
Department of Physics, Hong Kong University of Science and Technology, Clear Water Bay, Kowloon, Hong Kong SAR
}

\date{\today}
\begin{abstract}
There is a recent interest in studying odd elasticity in soft solids. Current focus has been on simple solids. However, many soft solids are structured and can exhibit nematic elasticity or viscoelasticity. Here we generalize the concept of odd elasticity to nematic elasticity. By rewriting the governing equation for two-dimensional nematic liquid crystals (LCs) in terms of complex Ginzburg--Landau equation, we propose an odd nematic elastic term and its stress term in the hydrodynamic model of nematic LCs. The odd nematic elasticity can be physically interpreted as non-reciprocal interactions between neighboring directors. In odd nematics we find that domain walls become self-propelled and are accompanied by a bidirectional flow, and point defects can self-spin, develop a spiral pattern, and induce a vortical flow. Interactions of a pair of defects show rich dynamics that are distinct from those in active nematics. As such, we have developed an odd general elasticity, which can be further generalized to other viscoelastic materials, and proposed a novel way to manipulate topological defects in nematic LCs. 
\end{abstract}


\maketitle

\textit{Introduction--}In conventional solid materials, the elasticity tensor is supposed to satisfy the Maxwell--Betti reciprocity relation, which allows for the existence of an elastic potential~\cite{truesdell1963remarks}. Recently, odd elasticity breaking such symmetry condition has been proposed theoretically~\cite{scheibner2020}. Realization of odd elasticity requires certain nonequilibrium process~\cite{fruchart2023odd}. For example, odd elasticity can emerge if an axial strain in a spring network can induce a transverse force~\cite{scheibner2020,braverman2021}. This type of odd material constant has been recently discovered in living systems comprising a monolayer of swimming starfish embryos~\cite{tan2022}, muscle fibres~\cite{shankar2024}, and dense human crowds~\cite{gu2025}. It can also be realized in mechanical metamaterials through certain active feedback loops~\cite{brandenbourger2019,veenstra2025}. 
In these odd elastic systems, net mechanical work can be extracted from cyclic processes, and they can also sustain elastic waves. Therefore, odd elasticity is promising for developing new mechano-intelligent active materials and devices. Indeed, a recent experiment has demonstrated that a ring of robots with odd elasticity can locomote on a variety of topographic surfaces, showing great adaptivity compared to conventional robots~\cite{veenstra2025adaptive}.

Many soft solids are structured, and thereby can encompass anisotropy (e.g. having liquid crystalline order) and viscoelasticity. Therefore, we ask: can generalized elasticity in structured soft solids, such as nematic elasticity in liquid crystals (LCs), also exhibit oddness? In this work, we theoretically propose a form of odd nematic elasticity, and study how this odd material property impacts the microstructure of a nematic LC. We further examine hydrodynamic effects of odd nematic elasticity. 


\begin{figure}[tbp]
	\centering
	\includegraphics[width=1\linewidth]{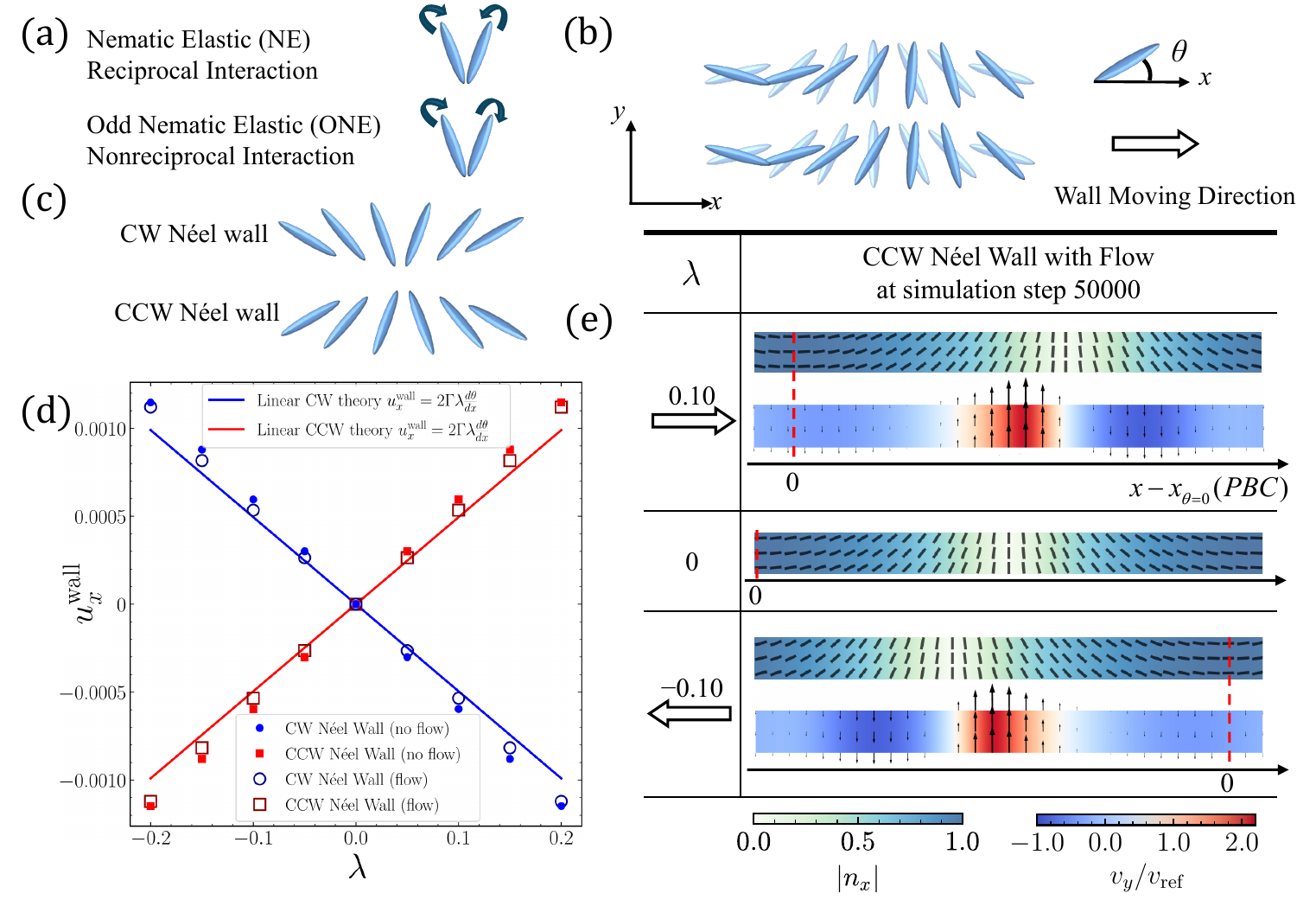}
	\vspace{\spaceBelowFigure}
	\phantomsubfloat{fig:Domain_Wall:a}
    \phantomsubfloat{fig:Domain_Wall:b}
    \phantomsubfloat{fig:Domain_Wall:c}
    \phantomsubfloat{fig:Domain_Wall:d}
    \phantomsubfloat{fig:Domain_Wall:e}
    \phantomsubfloat{fig:Domain_Wall:f}
    \vspace{-2.5\baselineskip}
    \caption{Locomotion of a N\'{e}el wall with odd nematic elasticity. 
    (a) Illustration of nematic elasticity (NE) and odd nematic elasticity (ONE) as reciprocal and non-reciprocal interactions between directors, respectively. 
    (b) Illustration of a domain wall self-propelled by ONE-induced collective rotation of directors. 
    (c) Schematic of clockwise (CW) and counter-clockwise (CCW) N\'{e}el. 
    (d) Self-propulsion velocities $u_x^{\mathrm{wall}}$ of CW and CCW N\'{e}el walls as functions of $\lambda$. Markers are simulation data with and without hydrodynamic effects, and solid lines are from Eq.~\ref{wall_speed}.
    (e) Director and the corresponding flow field of a CCW N\'{e}el wall with hydrodynamic effect. 
    Director orientations are indicated by black ticks, with the background color representing $|n_x|$. Arrows indicate flow direction, with color-coded intensity representing reduced velocity $v_y/v_{\mathrm{ref}}$.
    }
    \vspace{\spaceBelowCaption}
    \label{Domain_Wall}
\end{figure}

\textit{Model--}To construct odd nematic elasticity (ONE), we start with the two-dimensional (2D) Landau--de Gennes free energy density in tensorial form~\cite{de1993physics}:
\begin{equation}
\begin{aligned}
f=&-\frac{A}{4} \gamma Q_{i j} Q_{i j}+\frac{A}{4}\left(Q_{i j} Q_{i j}\right)^2+\frac{1}{2} L_1 Q_{i j, k} Q_{i j, k}\\&+\frac{1}{2} L_3 Q_{i j} Q_{k l, i} Q_{k l, j}, \label{LDG}
\end{aligned}
\end{equation}
where parameter $A$ sets the energy density scale, $\gamma$ controls the equilibrium scalar order parameter $S_0$ of the nematic, $L_1$ and $L_3$ are the elastic constants, and $Q_{i j}=S\left(n_i n_j-\frac{1}{2} \delta_{i j}\right)=\left(\begin{array}{cc}
Q_1 & Q_2 \\
Q_2 & -Q_1
\end{array}\right)$ is the symmetric and traceless tensorial order parameter with $S$ the scalar order parameter and $n_i$ the $i$'th component of the unit director. By introducing director angle $\theta$, we have $\hat{n}=(\cos \theta, \sin \theta)$. So $Q_1=\frac{S}{2} \cos 2 \theta$ and $Q_2=\frac{S}{2} \sin 2 \theta$. For the sake of conciseness, we set $L_3=0$ in the following theoretical discussion. 

The governing equation of the $\mathbf{Q}$-tensor is the Ginzburg--Landau equation $\dot{\mathbf{Q}}=-\Gamma \frac{\delta \mathcal{F}}{\delta \mathbf{Q}}$, where $\Gamma$ is a relaxation parameter and $\mathcal{F}=\int f\mathrm{d}V$ is the total free energy. This tensorial equation has been widely used to calculate the director field of an equilibrium nematic~\cite{ravnik2009landau}. By introducing a complex number $q \triangleq Q_1+\mathrm{i} Q_2$, the above tensorial equation can be recast to
\begin{equation}
\dot{q}=\Gamma\left(\frac{A}{2} \gamma q-2 A|q|^2 q+L_1 \nabla^2 q\right).
\end{equation}
By introducing two real parameters $\omega$ and $\lambda$, we convert the above equation into a Complex Ginzburg--Landau Equation (CGLE)~\cite{aranson2002}
\begin{equation}
\dot{q}=\left(\frac{\Gamma A \gamma}{2}+\mathrm{i} \omega\right) q-2 \Gamma A|q|^2 q+\Gamma(L_1+\mathrm{i} \lambda) \nabla^2 q, \label{CQE}
\end{equation}
where the first and third coefficients are complex. By introducing the 2D Levi-Civita tensor $\pmb{\varepsilon}=\left(\begin{array}{cc}
0 & 1 \\
-1 & 0
\end{array}\right)$, we can rewrite the above equation in the tensorial form:
\begin{equation}
\begin{aligned}
\dot{Q}_{i j}= &-\omega \varepsilon_{i k} Q_{k j}+\Gamma\left[\frac{A \gamma}{2} Q_{i j}-A\left(Q_{k l} Q_{l k}\right) Q_{i j}\right.\\
&\left.+L_1 \partial_k \partial_k Q_{i j}-\lambda \varepsilon_{i k} \partial_l \partial_l Q_{k j}\right]. \label{RQE}
\end{aligned}
\end{equation}
The equations with $L_3\neq0$ are included in~\cite{SM}. Assuming a uniform scalar order parameter $S$, the above equation can be reduced to
\begin{equation}
    \dot{\theta}=\frac{\omega}{2}+\Gamma (L_1 \nabla^2\theta - 2\lambda|\nabla\theta|^2),\label{eq_theta}
\end{equation}
from which we can see the physics of these complex coefficients: the $\omega$-term represents a spontaneous rotation of the director, and the $\lambda$-term drives director rotation wherever a gradient in the director angle $\theta$ is present.  
There is an interesting contrast between the $L_1$-term and the $\lambda$-term. Under nematic elasticity (NE) described by the $L_1$-term, two misaligned neighboring directors tend to rotate with opposite handedness to become aligned. This constitutes an effectively \textit{reciprocal} interaction [\cref{fig:Domain_Wall:a}]. In contrast, due to the $\lambda$-term, misaligned neighboring directors tend to rotate with the same handedness, resulting in an effectively \textit{non-reciprocal} interaction between them, breaking angular momentum conservation [\cref{fig:Domain_Wall:a}]. In this sense, we call the $\lambda$-term odd nematic elasticity (ONE). Similar non-reciprocal particle--particle interactions have been recently considered in several non-LC systems~\cite{chen2024, liu2024, fruchart2021non,saha2020scalar,rana2024,rana2024,osat2023non}. In what follows, we will focus on the effect of $\lambda$ in nematic LCs, but set $\omega=0$, the effect of which will be discussed in a separate work. 
To account for the hydrodynamic effects of ONE, an additional stress term $\pmb{\tau}^o=2\lambda (Q_1\nabla^2 Q_1+Q_2\nabla^2 Q_2)\pmb{\varepsilon}$ is included in the Beris--Edwards equation~\cite{SM}. The hydrodynamic model is numerically solved using a hybrid lattice Boltzmann method with parameters listed in below: $L_1=0.1$, $A=0.1$, $\Gamma=0.19$, $\gamma=0.1$, $\xi=0.8$, $\tau=500$.

\begin{figure}[tbp]
	\centering
\includegraphics[width=0.8\linewidth]{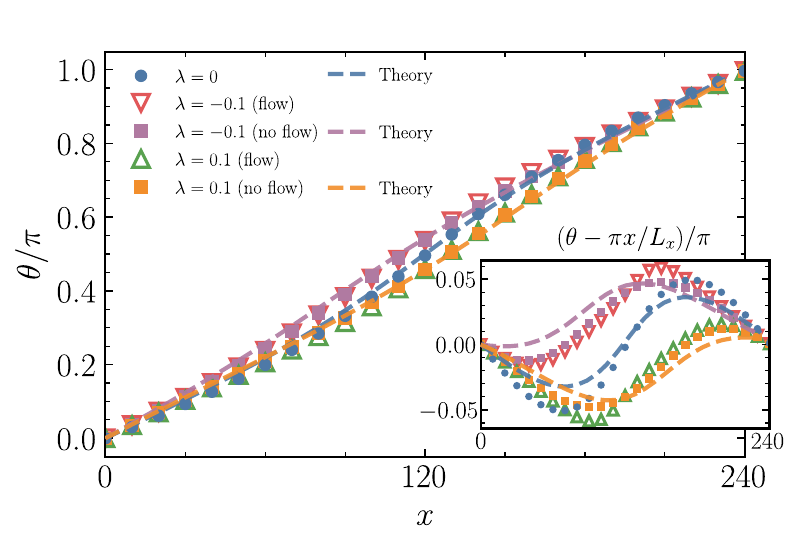}
	\vspace{\spaceBelowFigure}
    \vspace{-1.5\baselineskip}
    \caption{
    Director angle $\theta$ of a N\'{e}el wall with and without hydrodynamic effects. 
		}
    \vspace{\spaceBelowCaption}
    \label{fig:Wall_flow}
\end{figure}

\textit{Results--}According to the phase diagram of CGLEs, uniform state is stable in our odd nematic LC~\cite{aranson2002}. In what follows, we will examine the effects of ONE in different distorted states of the nematic.

We first study the dynamics of a domain wall. Specifically, we consider a one-dimensional (1D) splay-type N\'{e}el wall along the $y$-axis, with the director angle $\theta(x)$ as a function of $x$ varying from $\theta(x=-\infty)=0$ to $\theta(x=+\infty)=\pm \pi$ [\cref{fig:Domain_Wall:b,fig:Domain_Wall:c}]. 
Here we set $L_3=0.2$ to let the splay modulus be smaller than the bend modulus (i.e., $K_1<K_3$, see~\cite{SM}). According to Eq.~\ref{eq_theta}, the $\lambda$-term will drive all directors to rotate collectively in the presence of a nonzero spatial gradient of $\theta$ [\cref{fig:Domain_Wall:b}]. The governing equation of $\theta$ reads: $\dot{\theta}=\Gamma \left[ \left(L_1+\frac{L_3 S}{2} \cos 2\theta\right) \theta^{\prime \prime}-\left(\frac{L_3 S}{2} \sin 2\theta +2 \lambda\right) \left(\theta^{\prime}\right)^2 \right]$. The collective rotation of $\theta$ can lead to locomotion of the wall along the $x$-axis, breaking time-reversal symmetry. The actual locomotion direction depends on the wall's structure or handedness (clockwise or counter-clockwise, see \cref{fig:Domain_Wall:c}) as well as the sign of $\lambda$: for a counter-clockwise (clockwise) wall and positive (negative) $\lambda$, the wall will move in the $+x$ direction [video 1]; otherwise, it will move in the $-x$-direction [video 2]. 

By assuming no hydrodynamic effects and smallness in $\lambda$, we derive a simple formula for the locomotion speed~\cite{SM}: 
\begin{equation}
    u^\mathrm{wall}_x=\left.2 \Gamma \lambda \partial_x\theta\right|_{\text{wall}},\label{wall_speed}
\end{equation}
which implies that the wall's velocity is proportional to $\lambda$ and inversely depends on the wall width, characterized by $\partial_x \theta|_\mathrm{wall}$. Notably, the handedness of the wall is determined by the sign of $\partial_x \theta|_\mathrm{wall}$, and the earlier qualitative reasoning regarding the direction of wall's locomotion is consistent with this equation. This simple theoretical relation is further validated in our full simulations, showing good agreement, particularly for small $|\lambda|$, where hydrodynamic effects are negligible [\cref{fig:Domain_Wall:d}].

We further examine the hydrodynamic effects of ONE on a N\'{e}el wall. In 1D, the odd stress tensor simplifies to $\boldsymbol{\tau}^o=-2 \lambda S^2(\nabla \theta)^2 \boldsymbol{\varepsilon}$, which gives rise to a $y$-force $f^o_y=4 \lambda S^2 \theta^{\prime}(x) \theta^{\prime \prime}(x)$, driving spontaneous flow along the wall. Indeed, our full simulations reveal a bidirectional flow near the wall center [\cref{fig:Domain_Wall:e}]. Interestingly, we also find that this flow slightly reduces the wall's locomotion speed [\cref{fig:Domain_Wall:d}]. 

The above phenomena can be understood by analyzing the deformation of the N\'{e}el wall due to hydrodynamic effects. Focusing on the case $\lambda>0$, the profile of $\theta(x)$ is steeper than a linear function as a result of $K_1<K_3$ [\cref{fig:Wall_flow}]. This leads to $\theta''(x<0)>0$ and $\theta''(x>0)<0$, which, given that $\theta'(x)>0$, implies $f^o_y=4 \lambda S^2 \theta^{\prime}(x) \theta^{\prime \prime}(x)>0$ for $x<0$ and $f^o_y<0$ for $x>0$ [\cref{fig:Wall_flow}]. Because the spontaneous flow is driven by the $y$-force, the profiles of $f^o_y$ and $v_y$ are similar in shape, though $v_y$ appears smoother~\cite{SM}. The sign change in $f^o_y$ produces a bidirectional flow characterized by a strong shear ($\partial_x u_y>0$) near the center of the wall [\cref{fig:Domain_Wall:e}]. Consequently, the director tends to rotate more vertically due to flow-aligning effect, increasing $|\theta(x)|$. The flow also smooths the director field, slightly broadening the wall and reducing $\partial_x\theta$. Thus, the wall's locomotion is slowed down. 

Similar behaviors can also be found in bend walls~\cite{SM}. Moreover, the profiles of the odd force and flow for opposite-signed $\lambda$ are mirror images of each other~\cite{SM}. This symmetry feature is consistent with the fact that ONE breaks the mirror symmetry of the 2D system. In what follows, therefore, we will focus on $\lambda\ge0$.



\begin{figure}[thp]
	\centering
	\includegraphics[width=1\linewidth]{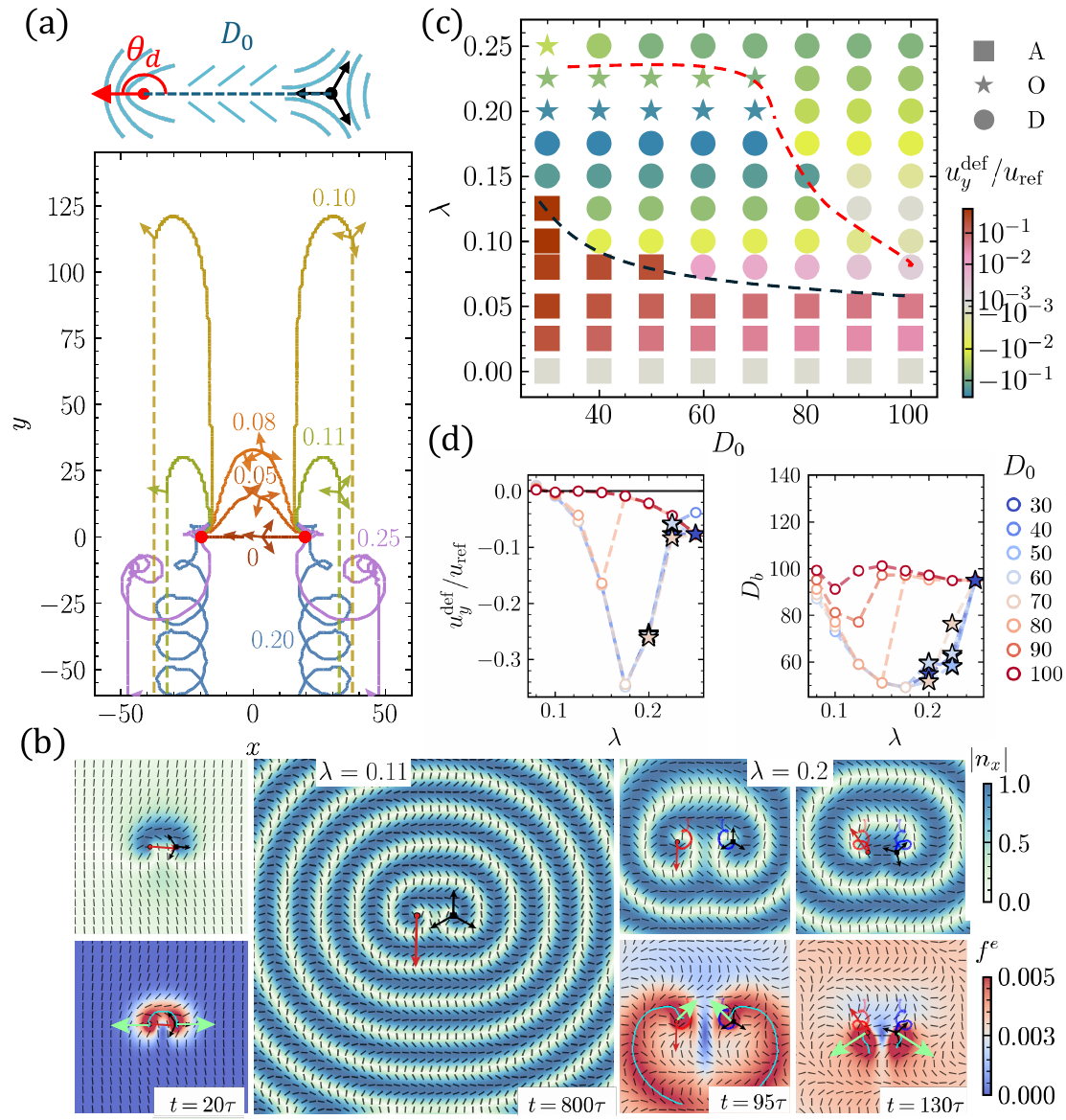}
	\vspace{\spaceBelowFigure}
	\phantomsubfloat{fig:Defect_Pair:a}
    \phantomsubfloat{fig:Defect_Pair:b}
    \phantomsubfloat{fig:Defect_Pair:c}
    \phantomsubfloat{fig:Defect_Pair:d}
    \vspace{-2.0\baselineskip}
    \caption{Dynamics of a pair of $\pm1/2$ defects without hydrodynamic effects. (a) Defect trajectories for initial separation $D_0=40$ for $0\le \lambda \le 0.25$.
    (b) Director field colored by $|n_x|$ (top) and by the total elastic energy density (bottom). Red and blue curves indicate the trajectories of $+1/2$ and $-1/2$ defects, respectively; the cyan skeleton marks the highly deformed domain wall, and light green arrows denote the its normal direction.
    (c) Phase diagram in the $(D_0,\lambda)$ plane, showing annihilation, oscillatory, and gliding regimes; colors indicate the averaged gliding velocity $u_y^\mathrm{def}/u_\mathrm{ref}$ evaluated at the equilibrium separation $D(t)=D(t_e)$, with $u_\mathrm{ref}=2\times10^{-3}$.
    (d) Dependence of the transverse velocity $v_y$ and the equilibrium separation $D(t_e)$ on $\lambda$.} 
    \vspace{\spaceBelowCaption}
    \label{Defect_Pair}
\end{figure}

ONE also impacts the behavior of topological defects.
For the sake of simplicity and without loss of generality, we adopt one-constant approximation by setting $L_3=0$.
Specifically, we consider a pair of $\pm1/2$ defects in an otherwise uniform nematic with an initial defect configuration characterized by $\theta_{d0}\equiv \theta_d(t=0)=\pi$ [i.e., the $+1/2$ defect points away from the $-1/2$ defect, see \cref{fig:Defect_Pair:a}]. We first simulate their dynamics without hydrodynamic effects. \cref{fig:Defect_Pair:a} shows their representative trajectories for different $\lambda$'s at a fixed initial separation distance $D_0=40$. In the absence of ONE, i.e., $\lambda=0$, the defect pair will approach each other along a straight line and annihilate at the midpoint. A small but finite ONE ($0 < \lambda < 0.1$) causes the defect pair to self-spin while drifting upward ($+y$ direction) with a prolonged annihilation time [\cref{fig:Defect_Pair:a}, video 3].
For a larger ONE ($\lambda \gtrsim 0.1$), defect dynamics become different: the defect pair first glides upward, then separates, and eventually turns away from each other and drift downward ($-y$ direction) at a constant speed, reaching a long-lived symmetric bound state~\cite{aranson1993theory} [video 4]. 
As $\lambda$ increases further to $0.2$ [\cref{fig:Defect_Pair:a}, video 5], the system enters oscillatory (O) mode as defects move in a curling path with a net downward drift. 
Upon increasing $\lambda$ to $0.25$, the defects roll along larger circular orbits and quickly enter the drift mode by moving steadily downward [\cref{fig:Defect_Pair:a}], which shows similar behavior for $0.1 \lesssim \lambda \lesssim 0.175$.
In the long-time limit (e.g., $t/\tau \ge 1600$, $\lambda=0.175$), all these bound states will gradually become unstable as defects will be pushed away~\cite{SM}. This long-time unstable behavior of the bound state was also observed in simulations of CGLEs~\cite{aranson1993formation}. 

The self-spinning of defects can be seen by examining an isolated defect of winding number $m$. We can assume the director angle $\theta=\theta_0(r)+m\phi$ in polar coordinates $(r,\phi)$, with $\theta_0(r)$ characterizing the radial variation of the director. For an undistorted defect, $\theta_0(r)\equiv\text{const.}$  Eq.~\ref{eq_theta} can be rewritten as $\dot{\theta}=\Gamma(L_1(\theta''_0+\theta'_0/r)-2\lambda|\theta'_0|^2-2\lambda m^2/r^2)$, 
the last two terms of which can drive a sustained clockwise rotation of the director field regardless of defect charge, and the closer to the defect core, the faster the rotation speed will be. As a result, circular-like domains will form from the defect core and propagate outward. The director field will eventually develop into a steady spiral pattern \cref{fig:Defect_Pair:b}, which is reminiscent of Frank--Read sources in solids and in nematic liquid crystals~\cite{long2024}, spiral waves driven by non-reciprocal interactions in living systems~\cite{rana2024defect,rana2024,liu2024}, and rotating spiral arms in homeotropically anchored nematics subjected to a rotating magnetic field~\cite{frisch1995}. A mathematical analysis of the spiral pattern formation has been discussed for CGLEs~\cite{aranson1993formation,aranson1993theory,aranson2002}.

The initial upward drift of defects can be intuitively understood by interpreting the director field between the defect pair in the early stage as a N\'{e}el wall, which should move to its right---i.e. upward [cf. \cref{fig:Domain_Wall:b}]. This initial upward motion is independent of the specific initial configuration of the defects~\cite{SM}. As the defects drift, the director field twists and gradually develops a spiral pattern. This effectively screens the elastic attraction between the defects, and a larger value of $\lambda$ allows the defects to drift a greater distance before annihilation. For $\lambda\gtrsim 0.1$, the defects no longer annihilate; instead, they undergo a steady or oscillatory downward drift. In the fully developed spiral pattern, the $y$-inversion symmetry is broken. The walls located below the defect pair are narrower and exhibit steeper director gradients, which in turn drag the defects downward [\cref{fig:Domain_Wall:b},~\cite{SM}].


\cref{fig:Defect_Pair:c} summarizes defect behaviors in a diagram as a function of $D_0$ and $\lambda$. The color encodes the drift velocity $u^\mathrm{def}_y$: for annihilation (A) mode, $u^\mathrm{def}_y$ is extracted by fitting the early-time motion of the defects for $t\approx 0$. For the drifting (D) and oscillatory (O) mode, $u^\mathrm{def}_y$ corresponds to the constant and average $y$-velocity reached in the bound state, respectively. 
A smaller $D_0$ indicates a stronger elastic attraction and thereby favors the A mode. A larger ONE coefficient $\lambda$ leads to a stronger odd force which favors the bound state, and thereby suppresses the A mode. 

In ~\cref{fig:Defect_Pair:d}, we show the dependence of $u^\mathrm{def}_y$ and the defect separation distance $D_b$ on $\lambda$ for different $D_0$'s in the bound state for O and D modes. Interestingly, data for the D mode in both figures can collapse onto two master curves. This implies that there are two possible bound states with different $D_b$ and $u^\mathrm{def}_y$. The ONE coefficient and the initial condition can dictate which bound state the system stays: for small $\lambda$'s and small $D_0$'s, the defect pair tends to stay closer and drift faster in their bound state. When the system transitions from one bound state to the other, it may enter the O mode with $D_b$ and $u^\mathrm{def}_y$ taking intermediate values~[\cref{fig:Defect_Pair:d}].
Although $u^\mathrm{def}$ and $D_b$ differ between the two bound states, they all collapse onto one master curve on which $|u^\mathrm{def}_y|$ decays as $D_b$ increases~\cite{SM}. The further apart the two defects are, the more axisymmetric they are, therefore the drift velocity will be smaller. 
\begin{figure}[tbp]
	\centering
	\includegraphics[width=1\linewidth]{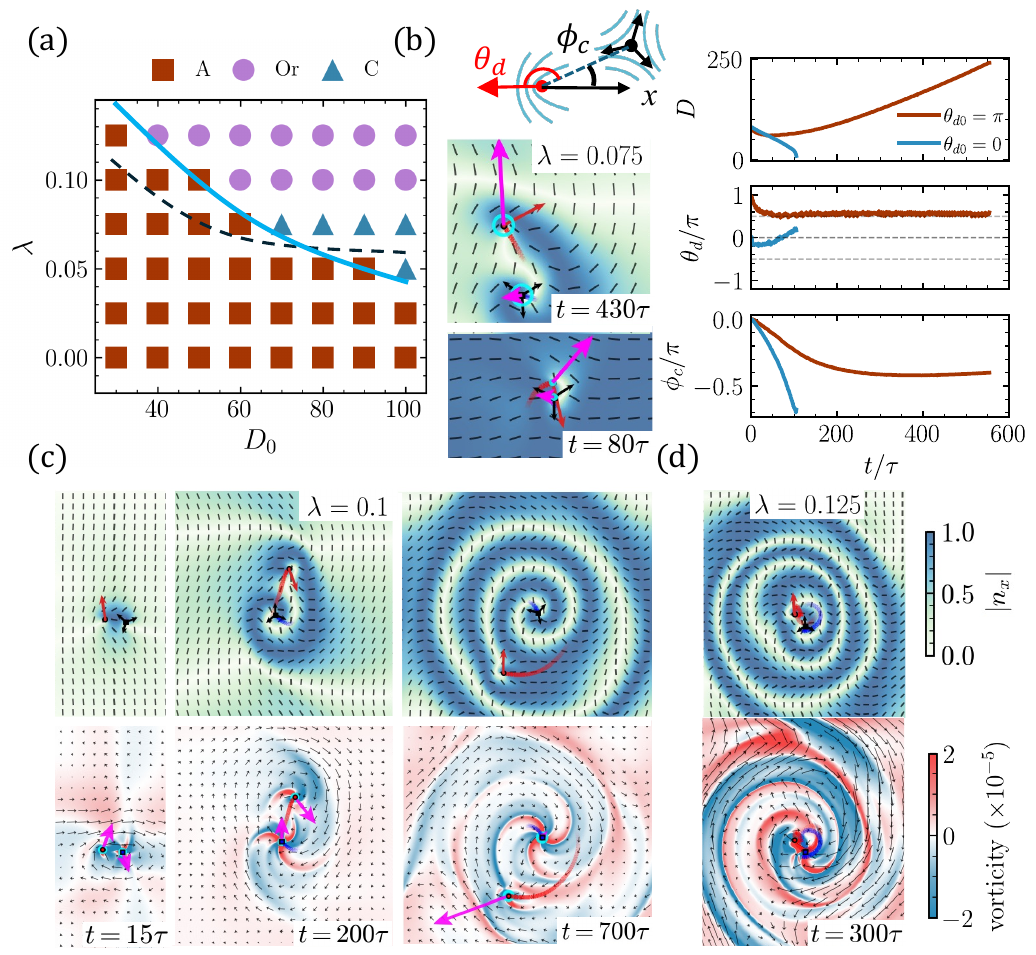}
	\vspace{\spaceBelowFigure}
    \phantomsubfloat{fig:Defect_Flow:a}
    \phantomsubfloat{fig:Defect_Flow:b}
    \phantomsubfloat{fig:Defect_Flow:c}
    \phantomsubfloat{fig:Defect_Flow:d}
    \phantomsubfloat{fig:Defect_Flow:e}
    \vspace{-2\baselineskip}
    \caption{
    Hydrodynamic effects on defect dynamics.
    (a) ``Phase'' diagram of a defect pair for $\theta_{d0}=\pi$ in the hydrodynamic simulation. The dashed black line denotes the phase boundary separating the annihilation (Ann.) phase from the other phases shown in \cref{Defect_Pair}, while the solid blue line indicates the new phase boundary.
    (b) Snapshots of the director field, colored by $|n_x|$, and the flow field, colored by the vorticity, at $D_0 = 80$ and $\lambda = 0.075$ for initial defect orientations $\theta_{d0} = \pi$ (top) and $\theta_{d0} = 0$ (bottom). Magenta arrows indicate the odd force $\mathbf{f}^{o}$ integrated around the defect cores, which are bounded by cyan circles. 
    The right panel shows time evolution of the defect separation $D$, the angle $\theta_d$ between the $+1/2$ defect orientation and the line connecting the $+1/2$ and $-1/2$ defects, and the angle $\phi_c$ of the connection line with initial defect orientations $\theta_{d0} = \pi$ and $0$, respectively.
    (c) Time evolution of the director and flow fields at $D_0 = 80$ and $\lambda = 0.1$.
    (d) Snapshots of the director and velocity fields illustrating the mutual orbital motion of the defects about their midpoint at $D_0 = 80$ and $\lambda = 0.125$.
 }
    \vspace{\spaceBelowCaption}
    \label{Defect_flow}
\end{figure}

The defect dynamics are qualitatively altered when hydrodynamic effects are considered. Three distinct regimes are identified in the simulation: annihilation (A), orbit (Or), and chase (C) mode [\cref{fig:Defect_Flow:a}]. In the A mode, defects neither annihilate at the midpoint, nor drift together in $\pm y$ directions persistently [\cite{SM}, \cref{fig:Defect_Flow:b}]---this behavior is substantially different from the A mode in the Ginzburg--Landau simulation. The Or mode is found in large $\lambda$ and $D_0$, in which the two defects will orbit around each other and also develop a spiral pattern. For intermediate $\lambda$ and large $D_0$, the defect pair enters the C mode as the $-1/2$ defect chases the $+1/2$ defect.

The hydrodynamic effects can be seen by considering an isolated, undistorted defect, around which the odd stress 
yields an odd force $\mathbf{f}^{o} = -\frac{4\lambda m^2S^{2}}{r^{3}}\hat{\boldsymbol{\phi}}$,
with $\hat{\boldsymbol{\phi}}$ the azimuthal unit vector. This force generates a clockwise vortex or swirling flow centered at the defect core. Therefore, the two defects will tend to orbit clockwise due to their hydrodynamic interactions---this explains the initial upward motion of $+1/2$ defect and downward motion of $-1/2$ defect [\cref{fig:Defect_Flow:c}]. Because of the presence of these vortex flows, the defect pair will not drift along the $y$-axis. The phase boundary between the A mode and other modes is slightly steeper than that in the Ginzburg--Landau model~[\cref{fig:Defect_Flow:a}]. Hydrodynamic effects generally accelerate defect annihilation. Therefore, the A mode is favored in the hydrodynamic model, particularly for small $D_0$. For large $D_0$, the system may enter the C mode, as the $+1/2$ defect is drifting, while the $-1/2$ defect follows via an elastic attraction force, forming an example of non-reciprocal interacting defects.

In the Or mode, the mutually orbiting defects are accompanied by strong circular flows generated by circular walls~[\cref{fig:Defect_Flow:d}], which can exert an azimuthal odd force $f_\phi^o=-4 \lambda S^2\left(\partial_r \theta\right)\left(\partial_r^2 \theta\right) \hat{\boldsymbol{\phi}}$. These circulating flow can stabilize the bound state by confining the defects. In this regime, more complex trajectories that mix orbital and chase behaviors are observed [video 10]. The time evolution of the separation distance $D$, the angle $\theta_d$ between the $+1/2$ defect orientation and the line connecting the two defects, and the angle $\phi_c$ of the connection line shows that the defects orbit at a constant speed with a fixed relative orientation, independent of the initial configuration~\cite{SM}. 

The initial orientations of the defects further influence the dynamics, particularly for intermediate values of $\lambda$ and during the early stages for larger $\lambda$. For example, at $D_0 = 80$, $\lambda = 0.075$, and $\theta_{d0} = \pi$, the $+1/2$ defect is rapidly dragged away by the odd force pointing away from the $-1/2$ defect, while the $-1/2$ defect chases the $+1/2$ defect [\cref{fig:Defect_Flow:b}]. 
However, if we set $\theta_{d0} = 0$ by rotating all directors by $\pi/2$, the $+1/2$ defect will move towards the $-1/2$ defect and eventually annihilate. This behavior is in stark contrast to the case without hydrodynamic effects, where a uniform rotation of the director field does not affect the steady-state defect dynamics~\cite{SM}. 

Note that hydrodynamic flow amplifies the asymmetry between the two defects. The director field around the two defects are different, the different anisotropic viscosities will give rise to different flows, which enhance their differences. In contrast, defect asymmetry is developed much slower in CGLEs~\cite{SM,aranson1993formation,aranson2002}.



\textit{Discussion--} Our work establishes a concrete physical system---odd nematic LCs---governed by the CGLE, in which the imaginary coefficient $\lambda$ is interpreted as an odd material constant (i.e. ONE). Additional physical effects not present in the original CGLE, such as elastic anisotropy and hydrodynamic effects, play important roles in shaping the dynamics of the odd nematic. In such odd nematic, domain walls locomote and point defects tend to self-spin and generate vortical flows---behaviors that stand in stark contrast to those observed in active nematics.
There also exists a deep connection between ONE and binary systems with non-reciprocal interactions. It can be seen from Eq.~\ref{CQE} that the coupling between $Q_1$ and $Q_2$ is inherently non-reciprocal. Existing studies have primarily focused on non-reciporcal interactions between conserved fields~\cite{liu2023non}. Note that odd viscosities in a uniform LC have been explored theoretically~\cite{souslov2020anisotropic}. It would be particularly interesting to investigate the interplay of a nonuniform director field and odd material properties. Future work may also consider odd elasticity and odd viscosity in more general settings\cite{walden2025odd}, extending the scope of odd matter beyond the framework presented here.

\textit{Acknowledgement--}We thank Shigeyuki Komura for helpful discussions. R.Z. acknowledges support from Hong Kong Research Grants Council. 

\bibliographystyle{jabbrv_apsrev4-2}
\bibliography{odd}
\end{document}